\pdfoutput=1
\documentclass[]{aastex631}
\usepackage{amsmath}
\usepackage{graphicx}
\usepackage{mathrsfs}
\usepackage{color}
\usepackage{url}
\usepackage{CJKutf8}
\usepackage{ulem}
\usepackage{wasysym}
\usepackage{pifont}

\received{2022}
\revised{\today}
\accepted{2022}
\shorttitle{Phaethon}
\shortauthors{Hui 2022}

\begin{document}

\title{
Perihelion Activity of (3200) Phaethon Is Not Dusty:
Evidence from STEREO/COR2 Observations
}

\correspondingauthor{Man-To Hui}
\email{mthui@must.edu.mo}

\author{\begin{CJK}{UTF8}{bsmi}Man-To Hui (許文韜)\end{CJK}}
\affiliation{State Key Laboratory of Lunar and Planetary Science, 
Macau University of Science and Technology, 
Avenida Wai Long, Taipa, Macau}
%\nocollaboration

%\nocollaboration

\begin{abstract}

We present an analysis of asteroid (3200) Phaethon using coronagraphic observations from 2008 to 2022 by the COR2 cameras onboard the twin Solar TErrestrial RElations Observatory (STEREO) spacecraft. Although undetected in individual images, Phaethon was visible in stacks combined from the same perihelion observations, yet only at small ($\la$30\degr) but not large ($\ga$150\degr) phase angles. The observations are in line with the contribution from a bare nucleus, thereby seriously contradicting the interpretation based on HI-1 observations that attributes the perihelion activity to the ejection of \micron-sized dust. We obtained an upper limit to the effective cross-section of \micron-sized dust to be $\la \! 10^{5}$ m$^{2}$, at least three orders of magnitude smaller than earlier estimates based on HI-1 data. On the contrary, the COR2 observations cannot rule out the existence of mm-sized or larger debris around Phaethon. However, the fact that no postperihelion debris tail has ever been detected for Phaethon suggests the unimportance of such dust in the perihelion activity. We thus conclude that the perihelion activity of Phaethon is highly unlikely relevant to the ejection of dust. Rather, we deduce that the activity is associated with gas emissions, possibly Fe I and/or Na D lines. To verify our conjecture and to fully understand the perihelion activity of Phaethon, more observations at small heliocentric distances are desired. We compile a list of observing windows ideal for the search of gas emissions of the asteroid from ground telescopes. The best opportunities will be during total solar eclipses.

\end{abstract}

\keywords{
asteroids: individual (Phaethon) --- methods: data analysis
}

\section{Introduction}
\label{sec_intro}

(3200) Phaethon is a near-Earth asteroid discovered by the Infrared Astronomical Satellite in 1983 \citep{1983IAUC.3878....1G}. Its current orbit is highly elongated with an orbital eccentricity of $e = 0.89$ and a semimajor axis of $a = 1.27$ au, resulting in a perihelion distance of merely $q = 0.14$ au. Phaethon is identified to be dynamically associated with the Geminid meteoroid stream \citep{1983IAUC.3881....1W,1993MNRAS.262..231W}, in contrast to most of other meteoroid streams, whose parent bodies are comets \citep{2006mspc.book.....J}. \citet{2006A&A...450L..25O,2008M&PSA..43.5055O} suggested that two additional asteroids (155140) 2005 UD and (225416) 1999 YC are potentially split fragments of Phaethon, together with Phaethon and the Geminids constituting the Phaethon-Geminid Complex (PGC). Various deep imaging searches for dust ejecta and gas emissions of Phaethon were carried out, returning only negative results \citep[][]{1996Icar..119..173C,1984Icar...59..296C,2005ApJ...624.1093H,2018AJ....156..238J,2019AJ....157..193J,2019AJ....158...30T,2008Icar..194..843W,2018ApJ...864L...9Y,2021PSJ.....2...23Y}. However, \citet{2010AJ....140.1519J}, \citet{2013ApJ...771L..36J}, \citet{2013AJ....145..154L}, and \citet{2017AJ....153...23H} reported that Phaethon underwent anomalous brightening and exhibited a tiny tail in the antisolar direction around its perihelion passages in 2009, 2012, and 2016 in observations taken by spacecraft Solar TErrestrial RElations Observatory (STEREO), likely due to the ejection of \micron-sized dust caused by thermal fracture. They all agreed that the observed perihelion activity is likely irrelevant to the Geminid meteoroid stream, because the estimated dust size is too small compared to the Geminids and the combined mass loss is also too small to sustain the stream effectively. Nonetheless, Phaethon is indisputably classified as an active asteroid \citep[and citations therein]{2022arXiv220301397J}. Besides, the dust trail along its orbit was recently observed at infrared and optical wavelengths \citep{2014AJ....148..135A,2020ApJS..246...64B,2022ApJ...936...81B}. Despite being consistent with the Geminid meteoroid stream, the dust trail slightly deviates from the orbit of Phaethon \citep{2022ApJ...936...81B}.

While the quest for the processes by which Phaethon produced the Geminid meteoroid stream (and other PGC members also) is by no means near the end, in this paper, we turn our primary focus on perihelion activity of Phaethon. The question whether the activity observed in HI-1 observations is attributed to the dust ejection remained unsettled, as the tail of Phaethon reminded us of the one of Mercury. While Phaethon was observed to possess a tiny tail around perihelion in images taken by the camera Heliospheric Imager-1 \citep[HI-1;][]{2009SoPh..254..387E} onboard STEREO \citep{2013ApJ...771L..36J,2017AJ....153...23H}, occasionally Mercury was seen to exhibit a similar albeit much brighter anti-sunward tail for reasons not yet fully understood \citep{2013PhDT.......602S}. One suggested hypothesis \citet{2013PhDT.......602S} is that the filter bandpasses of the HI-1 cameras have changed since launch, allowing for nontrivial transmission of sodium D-lines. Here, we investigate coronagraphic observations of Phaethon from STEREO, including those taken at never-before-seen large phase angles for the object, to characterise the nature of its perihelion activity. Our paper is organised in the following manner. We detail the coronagraphic observations and data reduction in Section \ref{sec_obs}, followed by an analysis and discussion of the observations in Section \ref{sec_nls}. A conjecture to explain the perihelion activity is put forward and future observing windows are identified in Section \ref{sec_conj}, and the summary is presented in Section \ref{sec_sum}.

\section{Observations \& Data Reduction}
\label{sec_obs}

\begin{figure}
\begin{center}
\gridline{\fig{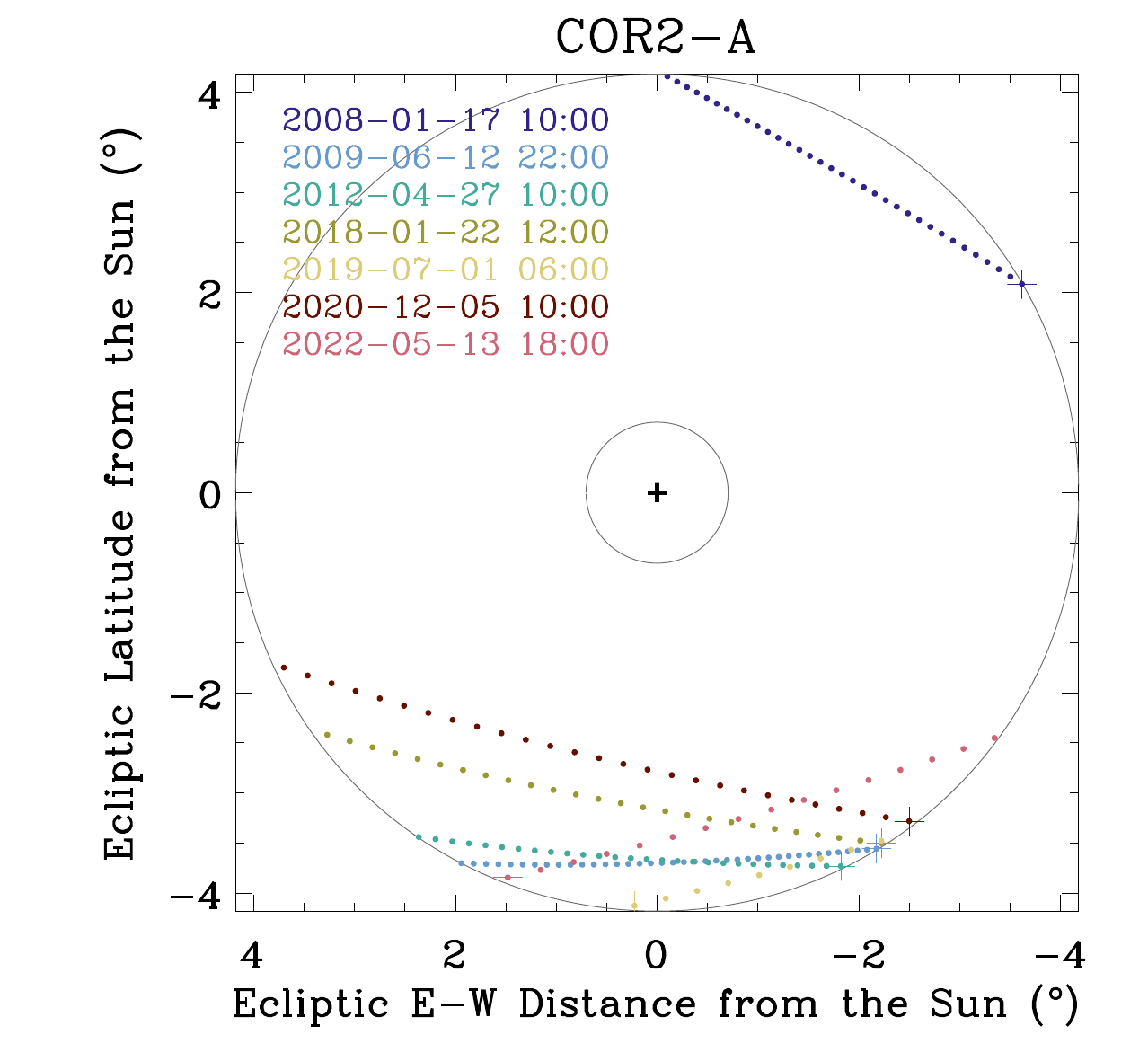}{0.5\textwidth}{(a)}
%}\gridline{
\fig{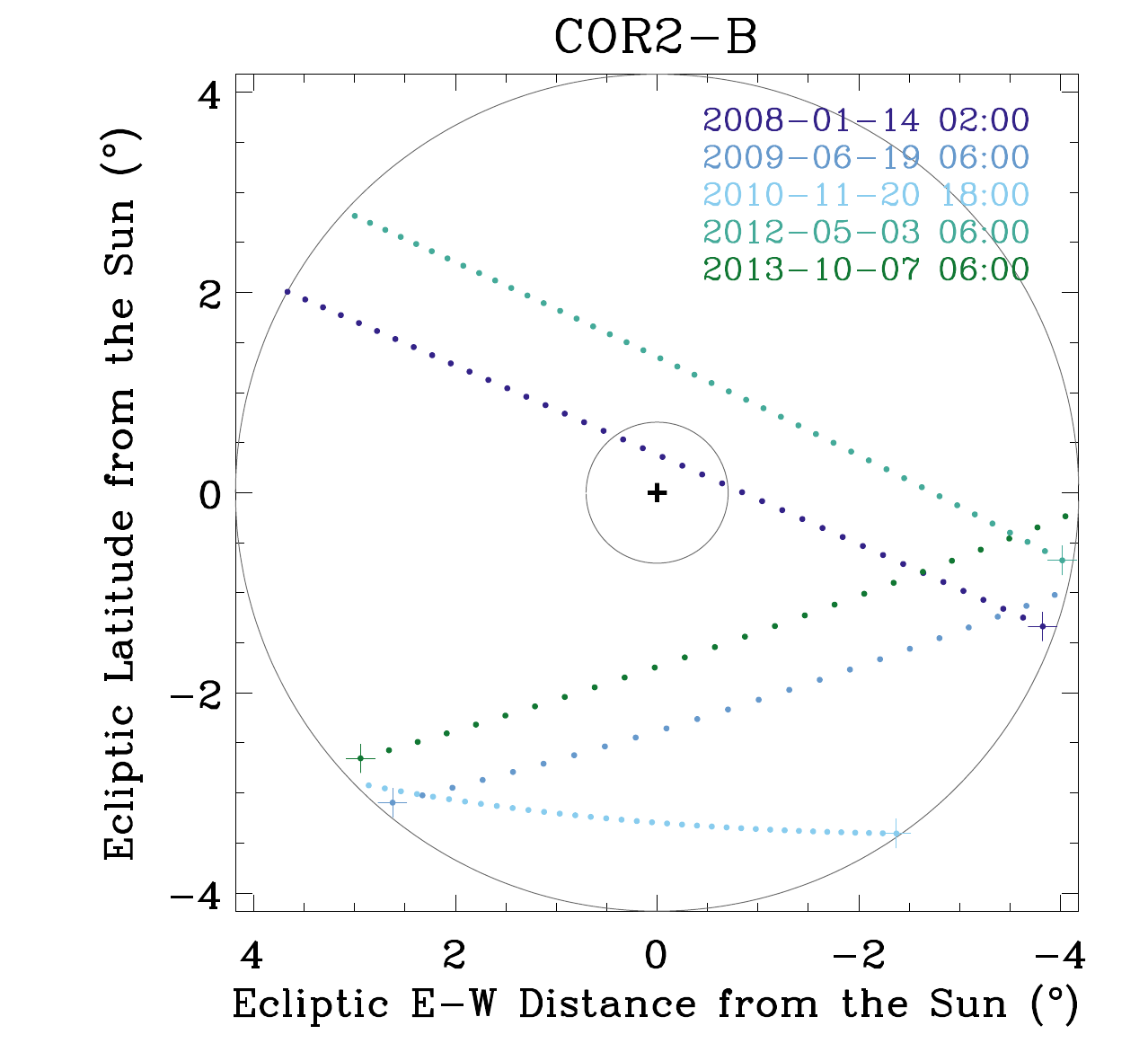}{0.5\textwidth}{(b)}}
\caption{
Transits of Phaethon in the annular FOVs (bounded by two concentric circles) of (a) COR2-A and (b) and COR2-B cameras. Trajectories from different apparitions are distinguished by different colours. The positions are plotted every two hours, with the first positions inside the FOVs indicated by symbol ``+" in the same colours. Corresponding epochs in UTC are given in the legends. The bold plus sign in black at the centre of either panel represents the centre of the Sun.
\label{fig:FOV}
} 
\end{center} 
\end{figure}

\begin{deluxetable*}{ccccccc}
%\tabletypesize{\scriptsize}
\tablecaption{Observing Geometry of Phaethon
%\rotate
\label{tab:vgeo}}
\tablewidth{0pt}
\tablehead{
Apparition & Date \& Time\tablenotemark{a} & Camera & Heliocentric Distance\tablenotemark{b} & Observer-centric Distance\tablenotemark{c} & Phase Angle\tablenotemark{d} & Visibility \\
& (UT) && $r_{\rm H}$ (au) & ${\it \Delta}$ (au) & $\alpha$ (\degr)
}
\startdata
2008 & Jan 17 10:38 -- Jan 20 03:38 & COR2-A & 0.206$\,\to\,$0.286 & [1.159, 1.242] & [14.0, 19.9] & \checkmark\\ 
     & Jan 14 01:08 -- Jan 17 08:08 & COR2-B & 0.140$\,\to\,$0.203 & [1.124, 1.192] & [4.4, 31.2] & \checkmark\\ %\hline
2009 & Jun 12 22:38 -- Jun 16 00:08 & COR2-A & 0.315$\,\to\,$0.225 & [1.168, 1.262] & [12.6, 17.8] & \checkmark\\
     & Jun 19 04:53 -- Jun 21 02:38 & COR2-B & 0.148$\,\overset{p}{\to}\,$0.144 & [0.912, 0.922] & [148.3, 162.7] & {\sffamily X}\\% \hline
2010 & Nov 20 18:24 -- Nov 23 18:24 & COR2-B & 0.245$\,\to\,$0.163 & [1.219, 1.309] & [17.3, 28.9] & \checkmark\\ %\hline
2012 & Apr 27 10:24 -- Apr 29 14:54 & COR2-A & 0.242$\,\to\,$0.180 & [1.121, 1.186] & [16.3, 22.5] & \checkmark\\
     & May 3 05:24 -- May 6 19:54 & COR2-B & 0.145$\,\to\,$0.231 & [1.125, 1.217] & [6.9, 29.9] & \checkmark\\ %\hline
2013 & Oct 7 04:24 -- Oct 9 06:24 & COR2-B & 0.143$\,\overset{p}{\to}\,$0.152 & [0.919, 0.933] & [147.5, 167.6] & {\sffamily X}\\% \hline
2018 & Jan 22 12:24 -- Jan 24 14:54 & COR2-A & 0.184$\,\to\,$0.143 & [1.082, 1.126] & [18.9, 28.8] & \checkmark\\ %\hline
2019 & Jul 1 05:39 -- Jul 1 21:54 & COR2-A & 0.159$\,\to\,$0.148 & [0.819, 0.832] & [152.0, 154.4] & {\sffamily X}\\% \hline
2020 & Dec 5 10:24 -- Dec 7 14:39 & COR2-A & 0.168$\,\overset{p}{\to}\,$0.140 & [1.079, 1.111] & [17.8, 29.7] & \checkmark\\% \hline
2022 & May 13 17:39 -- May 14 20:54 & COR2-A & 0.153$\,\to\,$0.141 & [0.824, 0.836] & [152.4, 157.1] & {\sffamily X}\\
\enddata
\tablenotetext{a}{Timestamps of the first and last available images in which Phaethon was apparently inside the outer edge of the camera's FOV.}
\tablenotetext{b}{The first and last numbers respectively correspond to the heliocentric distances of Phaethon in the first and last available images. If the perihelion passage was covered in the FOV, a letter ``p" is inserted in between above the arrow.}
\tablenotetext{c}{The interval gives the observed minimum and maximum observer-centric distances, not necessarily corresponding to the distances of Phaethon in the first and last available images from the apparition.}
\tablenotetext{d}{The angle of Sun-Phaethon-observer. The interval simply shows the observed minimum and maximum phase angles.}
%\tablecomments{}
\end{deluxetable*}

\begin{figure}
\epsscale{1.0}
\begin{center}
\plotone{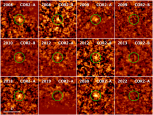}
\caption{
Phaethon in COR2 images around different perihelion returns. The images are median stacks from individual images taken by the same cameras in the same apparitions with alignment on the ephemeris motion of Phaethon (marked by a green circle in each panel). We are unable to confidently detect Phaethon in the observations taken at large phase angles (COR2-B in the apparitions of 2009 and 2013, COR2-A in 2019 and 2022). J2000 equatorial north is up and east is left. A scale bar of 5\arcmin~in length applicable to all of the panels is shown in the lower left corner.
\label{fig:obs}
} 
\end{center} 
\end{figure}

STEREO \citep{2008SSRv..136....5K} consists of twin solar probes moving in Earth-like heliocentric orbits yet in opposite directions with respect to Earth. The one leading Earth is STEREO-A, and the other one is STEREO-B. Both of the spacecraft carry the Sun Earth Connection Coronal and Heliospheric Investigation (SECCHI) onboard, which is a suite of five different telescopes, including an extreme ultraviolet imager EUVI, two Lyot coronagraphs COR1 and COR2, and two heliospheric imagers HI-1 and HI-2 \citep{2008SSRv..136...67H}. Since 2014 October, communication with STEREO-B has been lost due to hardware anomalies.\footnote{\url{https://stereo-ssc.nascom.nasa.gov/behind_status.shtml}} In our paper, only COR2 and HI-1 are the relevant telescopes, and the others will be ignored.

Phaethon has been studied using observations taken by both HI-1 cameras \citep{2010AJ....140.1519J,2013AJ....145..154L,2013ApJ...771L..36J,2017AJ....153...23H}. These images have an angular resolution of $\sim$70\arcsec~pixel$^{-1}$ and covered a square field of view (FOV) of $\sim\!20\degr \times 20\degr$ along the ecliptic at solar elongations between $\sim$4\degr~and 24\degr~\citep{2009SoPh..254..387E}. On the contrary, observations of Phaethon in the COR2 cameras onboard STEREO-A and STEREO-B (labeled COR2-A and COR2-B, respectively) have never been previously reported. Both cameras monitor an annular region of $\sim$2.5-15 $R_{\odot}$ (where $R_{\odot}$ is the apparent solar radius at a heliocentric distance of 1 au, or $\sim$0\fdg7-4\fdg0) with the Sun blocked at the centre. The effective bandpasses of the cameras are both $\sim$650-750 nm \citep{2008SSRv..136...67H}.

Using JPL Horizons' solution to the orbit of Phaethon, we identified that the object has also transited the FOVs of COR2-A and COR2-B around perihelion for multiple times since the spacecraft became operational, including four apparitions at large phase angles ($\alpha \ga 150\degr$), placing interplanetary dust in a regime where strong forward-scattering enhancements may occur \citep{2004come.book..577K,2007ICQ....29...39M}. We show the apparent trajectories of Phaethon within the FOVs of COR2-A and COR2-B in Figure \ref{fig:FOV} and tabulate the observing geometry in Table \ref{tab:vgeo}. Accordingly, we collected the COR2 total brightness images around the time of the transits of Phaethon. All of these level-0.5 Flexible Image Transport System (FITS) images are $2048 \times 2048$ pixels in size in an unbinned mode and have a pixel scale of 14\farcs7 and an individual exposure of 6 seconds in duration. 

The images were first calibrated by bias subtraction and vignetting correction and then normalised by the exposure time in {\tt IDL} using {\tt SolarSoftWare} \citep[{\tt SSW};][]{1998SoPh..182..497F}. Next, in order to maximally suppress the corona, we computed a median background for each calibrated image from neighbouring individual exposures, which was subsequently subtracted from the calibrated image. The resulting images are basically free from artifacts, in which background stars are clearly visible. Only slight residuals of time-varying fine features of the corona remained near the inner edge of the annular FOV in some of the images, but they posed no strong influence on our study of Phaethon on most occasions. Despite that the astrometric information is available in the FITS file headers, we noticed that the accuracy tended to decline towards the edge of the images due to the field distortion, as catalogued and observed field stars thereabouts showed visible positional discrepancies of a few pixels at worst. Thus, astrometric calibration of the images was performed with the software package {\tt astrometry.net} \citep{2010AJ....139.1782L}, resulting in improved solutions via visual inspection.

We attempted to search for Phaethon by visually examining individual COR2 images in which Phaethon was in the FOV, but we were unable to robustly detect it above the noise level around the ephemeris positions. In order to maximally suppress noise, we combined all of these images taken by the same cameras in the same apparitions with respect to the calculated apparent motion of Phaethon. We thereby managed to immediately identify the target at the exact calculated ephemeris positions in the stacks obtained at small phase angles $\la \! 30\degr$, as it was basically the brightest object therein. The FWHM of Phaethon was measured to be consistent with those of field stars in individual images, $\sim$2-3 pixels. On the contrary, we were unable to spot Phaethon in the stacks combined from images taken at large phase angles $\ga \! 150\degr$. In Figure \ref{fig:obs} we show the combined images of Phaethon, including the nondetections. The visibility is summarised both in Tables \ref{tab:vgeo} and \ref{tab:phot} for convenience.

One may notice an observational bias from Table \ref{tab:phot} that the stacks which show Phaethon were combined from systematically more individual images than those that do not. To get rid of the bias, we evenly divided individual images taken by the same cameras from the same perihelion passages in which Phaethon was visible in the final stacks into two groups, if the total numbers of images covering Phaethon are less than 200, otherwise three groups. Then, we still combined individual images from each group into a new stack with alignment on the ephemeris motion of Phaethon. As a result, we were still able to see Phaethon at the predicted positions in all but three (one from 2008 by COR2-A, and the two others from 2012 by COR2-B) of the new stacks. The three stacks showing no hint of Phaethon were combined from individual exposures where Phaethon was $\ga$0.2 au from the Sun, further than in any other images. Therefore, we can firmly rule out the possibility that the visibility of Phaethon is caused by the observational bias. Rather, it indicates the authentic apparent brightness of the asteroid.

\begin{deluxetable*}{ccc|c|r|r|r|c|c|c}
%\tabletypesize{\scriptsize}
\tablecaption{Summary of Photometric Calibration and Observations of Phaethon
\rotate
\label{tab:phot}}
\tablewidth{0pt}
\tablehead{
&&& \multicolumn{4}{c|}{Observations of Phaethon} &
\multicolumn{3}{c}{Best-fit Photometric Parameters} \\ \cline{4-10}
Apparition & Camera & Number & Visibility & \multicolumn{1}{c|}{App. Mag.\tablenotemark{b}} & \multicolumn{1}{c|}{Red. Mag.\tablenotemark{c}} & \multicolumn{1}{c|}{Cross-section\tablenotemark{d}} & Zero-point & 
\multicolumn{2}{c}{Linear Coefficients} \\ \cline{9-10} %\cline{11-12}
&& \multicolumn{1}{c|}{of Images\tablenotemark{a}} & & \multicolumn{1}{c|}{$\bar{m}$} & \multicolumn{1}{c|}{$\bar{H} \left(\bar{\alpha} \right)$} & \multicolumn{1}{c|}{$\bar{{\it \Xi}}_{\rm d}$ (m$^2$)} & $m_0$ & $\mathcal{C}_1$ & $\mathcal{C}_3$
}
\startdata
2008 & COR2-A & 186 (131)
                 & \checkmark & $11.17 \pm 0.22$
                 & $13.85 \pm 0.22$
                 & $\left(+1.6 \pm 3.8\right) \times 10^{7}$
                 & $12.81 \pm 0.12$
                 & $-0.261 \pm 0.019$
                 & $+0.199 \pm 0.026$ \\ 
     & COR2-B & 222 (120)
                 & \checkmark & $10.83 \pm 0.27$
                 & $14.46 \pm 0.27$
                 & $\left(-1.7 \pm 3.3 \right) \times 10^7$
                 & $12.67 \pm 0.12$
                 & $-0.277 \pm 0.013$
                 & $+0.217 \pm 0.014$ \\ %\hline
2009 & COR2-A & 250 (148)
                 & \checkmark & $11.48 \pm 0.22$ 
                 & $13.94 \pm 0.22$
                 & $\left(+5.9 \pm 34.8\right) \times 10^6$
                 & $12.98 \pm 0.15$
                 & $-0.249 \pm 0.031$
                 & $+0.219 \pm 0.026$ \\
     & COR2-B & 144 (94)
                 & {\sffamily X} & $\ge 12.30 \pm 0.18$ 
                 & $\ge 16.74 \pm 0.18$
                 & $\le \left(+9.5 \pm 10.5\right) \times 10^{4}$
                 & $12.93 \pm 0.13$
                 & $-0.236 \pm 0.021$
                 & $+0.212 \pm 0.023$ \\% \hline
2010 & COR2-B & 282 (217)
                 & \checkmark & $11.60 \pm 0.25$ 
                 & $14.61 \pm 0.25$
                 & $\left(-6.5 \pm 24.8\right) \times 10^6$
                 & $12.84 \pm 0.12$
                 & $-0.256 \pm 0.023$
                 & $+0.229 \pm 0.021$ \\ %\hline
2012 & COR2-A & 208 (157)
                 & \checkmark & $11.73 \pm 0.25$ 
                 & $14.83 \pm 0.25$
                 & $\left(-2.3 \pm 3.9\right) \times 10^7$
                 & $12.96 \pm 0.10$
                 & $-0.256 \pm 0.021$
                 & $+0.227 \pm 0.021$ \\
     & COR2-B & 282 (261)
                 & \checkmark & $11.17 \pm 0.30$ 
                 & $14.58 \pm 0.30$
                 & $\left(-1.6 \pm 3.2\right) \times 10^7$
                 & $12.89 \pm 0.09$
                 & $-0.280 \pm 0.017$
                 & $+0.263 \pm 0.023$ \\ %\hline
2013 & COR2-B & 136 (95)
                 & {\sffamily X} & $\ge 12.37 \pm 0.18$ 
                 & $\ge 16.77 \pm 0.18$
                 & $\le \left(+5.3 \pm 5.8\right) \times 10^{4}$
                 & $13.15 \pm 0.14$
                 & $-0.279 \pm 0.024$
                 & $+0.298 \pm 0.027$ \\% \hline
2018 & COR2-A & 198 (141)
                 & \checkmark & $11.01 \pm 0.21$ 
                 & $14.78 \pm 0.21$
                 & $\left(-1.6 \pm 3.4\right) \times 10^7$
                 & $12.88 \pm 0.15$
                 & $-0.259 \pm 0.033$
                 & $+0.220 \pm 0.039$ \\ %\hline
2019 & COR2-A & 93 (50)
                 & {\sffamily X} & $\ge 12.49 \pm 0.15$ 
                 & $\ge 16.99 \pm 0.15$
                 & $\le \left(+1.6 \pm 1.7\right) \times 10^{5}$
                 & $12.96 \pm 0.15$
                 & $-0.241 \pm 0.032$
                 & $+0.217 \pm 0.039$ \\% \hline
2020 & COR2-A & 213 (160)
                 & \checkmark & $10.77 \pm 0.21$ 
                 & $14.70 \pm 0.21$
                 & $\left(-1.4 \pm 3.3\right) \times 10^7$
                 & $13.01 \pm 0.12$
                 & $-0.228 \pm 0.026$
                 & $+0.208 \pm 0.028$ \\ %\hline
2022 & COR2-A & 155 (84)
                 & {\sffamily X} & $\ge 12.49 \pm 0.15$ 
                 & $\ge 17.09 \pm 0.15$
                 & $\le \left(+1.2 \pm 1.3\right) \times 10^{5}$
                 & $13.06 \pm 0.11$
                 & $-0.254 \pm 0.029$
                 & $+0.239 \pm 0.031$ \\ \hline
Overall & COR2-A & 1303 
                 & \multicolumn{4}{c|}{N/A}
                 & $12.96 \pm 0.15$
                 & $-0.251 \pm 0.029$
                 & $+0.219 \pm 0.031$ \\
        & COR2-B & 1066
                 & \multicolumn{4}{c|}{N/A}
                 & $12.88 \pm 0.18$
                 & $-0.271 \pm 0.024$
                 & $+0.246 \pm 0.036$ \\
\enddata
\tablenotetext{a}{Numbers of images used for photometric calibration (unbracketed) and combined for Phaethon (bracketed).}
\tablenotetext{b}{Apparent magnitude of Phaethon in the COR2 bandpasses.}
\tablenotetext{c}{Reduced (distance-normalised) magnitude of Phaethon in the COR2 bandpasses.}
\tablenotetext{d}{Total effective scattering cross-section of ejected \micron-sized dust.}
\tablecomments{
Parameters $\mathcal{C}_{j}$ ($j=1,2,3$) correspond to the linear coefficients for the $g$, $r$, and $i$-band star magnitudes in the Pan-STARRS photometric system. We set $\mathcal{C}_{2} \equiv 1$ in the Levenberg-Marquardt optimisation. The reported values of the image zero-points and linear coefficients and the associated errors are weighted means and standard deviations, respectively. Values in photometry of Phaethon preceded by inequality signs are $3\sigma$ limits.}
\end{deluxetable*}

\section{Analysis}
\label{sec_nls}

\subsection{Photometry}
\label{ssec_phot}

In order to convert the observed flux (in DN s$^{-1}$) to the apparent magnitude of Phaethon, accurate image zero-points of COR2-A and COR2-B would be needed. However, there were only preliminary results from pre-flight tests by the STEREO team. Therefore, we had to determine the image zero-points by performing photometry of field stars in individual images. The apparent magnitude and the observed flux of a star (labeled by the symbol ``$\ast$") in a COR2 image, denoted as $m_{\ast}$ and $F_{\ast}$, respectively, are related to the image zero-point $m_{0}$ by
\begin{equation}
m_{\ast} = \underbrace{-2.5 \log F_{\ast}}_{\widetilde{m}_{\ast}} + \,m_{0}
\label{eq_mstar},
\end{equation}
\noindent in which $\widetilde{m}_{\ast} = -2.5 \log F_{\ast}$ is the instrumental magnitude of the star. Since the measured pre-flight effective bandpasses of the COR2 cameras are different from any photometric standard, we adopted the colour mixing method in \citet{2010SoPh..264..433B} and approximated the apparent magnitude of some star in the COR2 bandpass as a linear combination of the apparent magnitudes of the star in three different photometric standard bandpasses, i.e.,
\begin{equation}
m_{\ast} = \sum_{j=1}^{3} \mathcal{C}_j m_{\ast,j}
\label{eq_mstar2},
\end{equation}
\noindent where $\mathcal{C}_{j}$ ($j=1,2,3$) is the linear coefficient in the $j$-th band. With multiple field stars, the linear coefficients and the image zero-point can be determined by minimising  the goodness of fit
\begin{equation}
\chi^2 = \sum_{k} \frac{\left[\widetilde{m}_{\ast_{k}} + m_{0} - \sum_{j=1}^{3} \mathcal{C}_{j} m_{\ast_{k},j}\right]^2}{\left(\Delta{\widetilde{m}_{\ast_{k}}}\right)^2 + \sum_{j=1}^{3} \left(\mathcal{C}_{j} \Delta{m_{\ast_{k},j}}\right)^2}
\label{eq_chi_sqr}.
\end{equation}
\noindent Here, $\Delta {\widetilde{m}_{\ast_{k}}}$ and $\Delta {m_{\ast_{k},j}}$ are the uncertainties in the instrumental magnitude and the apparent magnitude in the $j$-th band, respectively, of the $k$-th star ($\ast_{k}$) in the image. 

The observed fluxes of field stars were measured using a circular aperture of 3 pixels in radius, whose uncertainties were obtained by propagating errors from the Poisson statistics and fluctuation in the sky background measured from a concentric annular region having radii between 1.5$\times$ and 2.5$\times$ the aperture radius. Given the effective bandpasses of the COR2 cameras, we set $\mathcal{C}_{j}$ ($j=1,2,3$) respectively correspond to the linear coefficients for $g$, $r$, and $i$-band star magnitudes in the Pan-STARRS photometric system from the ATLAS All-Sky Stellar Reference Catalog \citep[Refcat2;][]{2018ApJ...867..105T} for photometric calibration. The Levenberg-Marquardt optimisation routine {\tt MPFIT} \citep{2009ASPC..411..251M} was exploited to obtain the best-fit photometric parameters along with their uncertainties for each of the COR2 images. During initial tests, we found that {\tt MPFIT} would converge to different local minima with slightly different initial guesses for the parameters to be solved and yet their $\chi^2$ values were comparable. Also given the fact that the COR2 bandpasses largely overlap with the $r$ band in the Pan-STARRS photometric system, we thus fixed $\mathcal{C}_2 \equiv 1$ and performed the optimisation for other parameters. Stars fainter than $\sim$9.5 mag in the three bands and those with magnitudes of observed-minus-calculated residuals over $3\sigma$ were discarded. 

\begin{figure}
\epsscale{0.8}
\begin{center}
\plotone{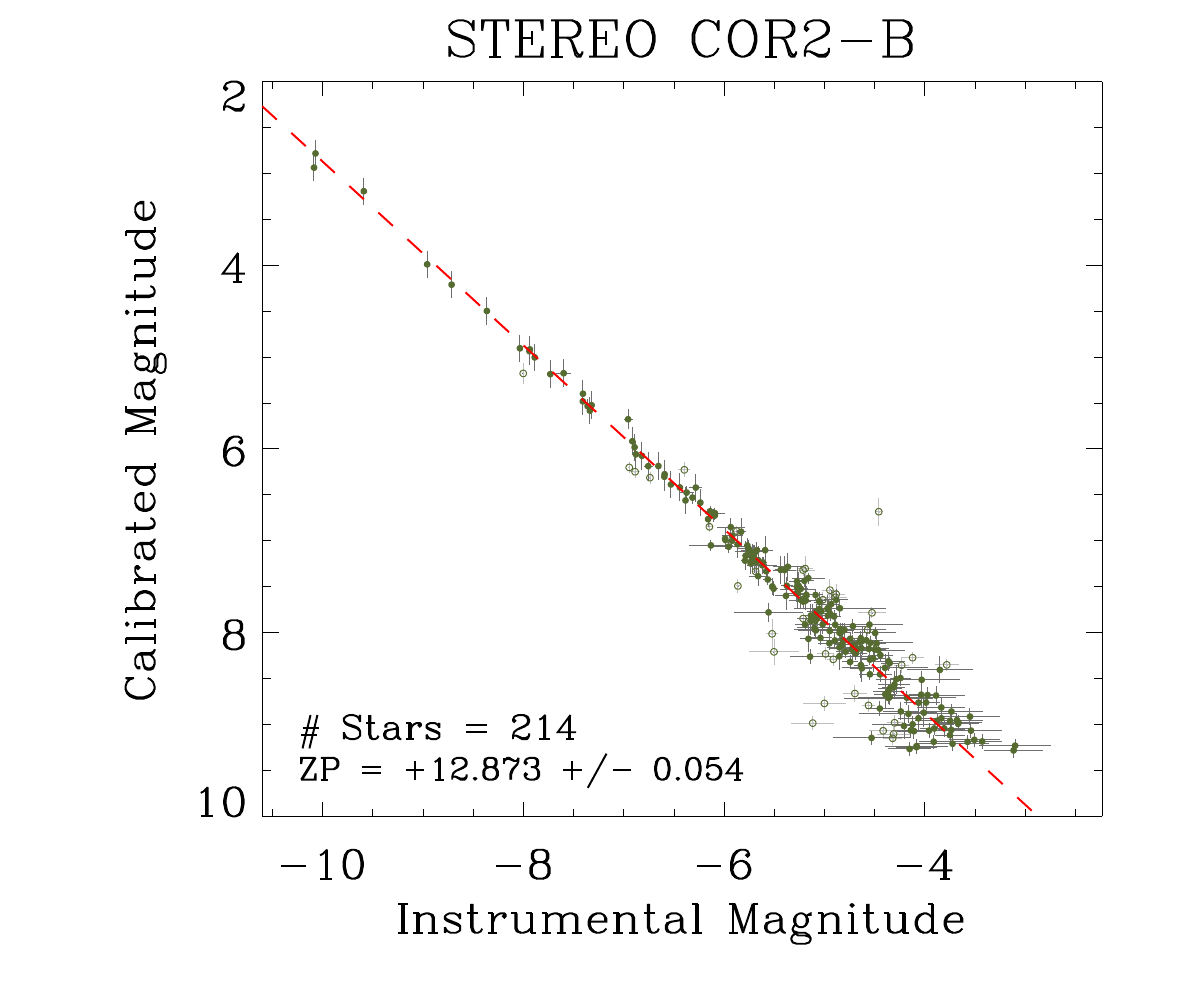}
\caption{
Example of fitting the photometric parameters for a COR2-B image from 2012 May 3. Field stars used and rejected in the Levenberg-Marquardt optimisation are plotted as olive filled and open circles, respectively. The best-fit function is shown as the red dashed line. In the lower left corner, we present the number of used stars and the image zero-point (abbreviated as ``ZP" therein).
\label{fig:calimg}
} 
\end{center} 
\end{figure}

\begin{figure}
\epsscale{0.8}
\begin{center}
\plotone{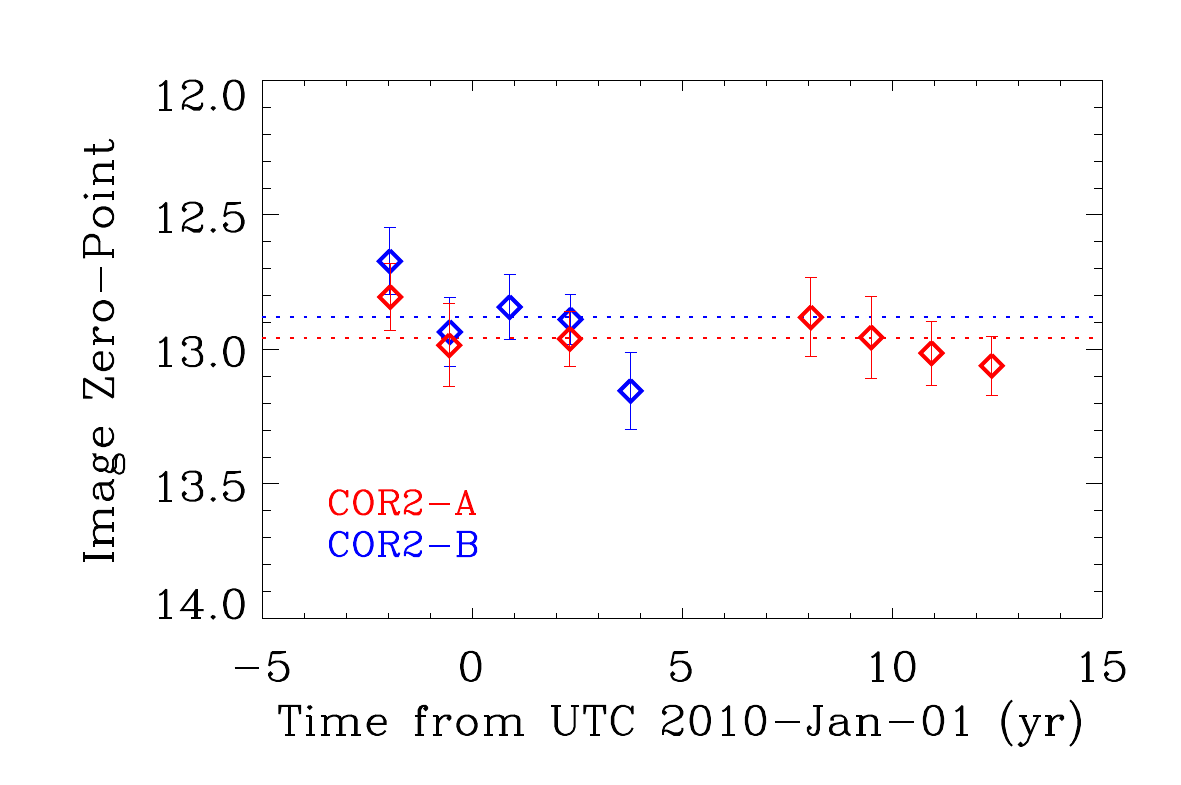}
\caption{
Zero-points of COR2 images in their respective bandpasses versus time. Data points of COR2-A and COR2-B are plotted in red and blue, respectively. The dotted lines are weighted mean zero-point values of the two cameras. There is tentative evidence that the zero-points have been changing over time, however, the uncertainties are too large for further interpretation.
\label{fig:ZP_ageing}
} 
\end{center} 
\end{figure}

Figure \ref{fig:calimg} shows a typical example of the best-fit results for a COR2 image and the columns under ``Best-fit Photometric Parameters" in Table \ref{tab:phot} are a summary of our best-fit results of the photometric calibration. We have verified that the best-fit results are robust, as adjusting settings such as the size of the photometric aperture and the cutoff threshold for outlier rejection would not alter the best-fit results over their respective uncertainties whatsoever. We plot the best-fit image zero-points in Figure \ref{fig:ZP_ageing} to investigate if there is an ageing effect that causes a drift therein over the course of the past multiple perihelion returns of Phaethon. Although we can spot some temporal variations, they are not significantly greater than the uncertainties. We repeated the aforementioned procedures yet with slightly larger circular apertures for photometry, finding that the results remained unchanged within the respective uncertainty levels. Therefore, we conclude that the ageing effect of the COR2 cameras is negligible in comparison to the uncertainties in photometric measurements. We calculated the weighted mean and standard deviation values of the photometric parameters for all of the COR2-A and COR2-B images separately (Table \ref{tab:phot}), which we adopted to obtain the apparent magnitude of Phaethon (see the column ``App. Mag." in Table \ref{tab:phot}). In cases of the nondetections, $3\sigma$ lower limits are given instead.

\begin{figure}
\epsscale{1.0}
\begin{center}
\plotone{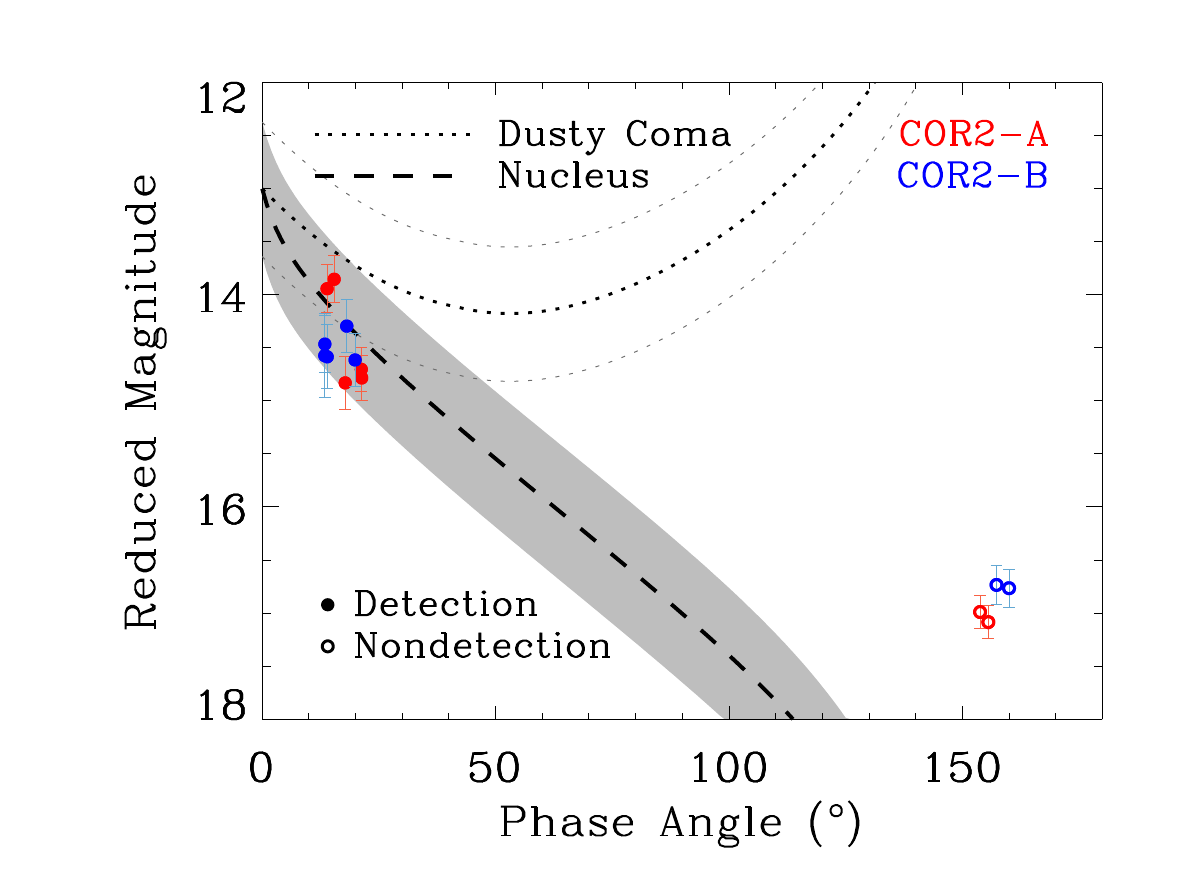}
\caption{
Reduced magnitude of Phaethon versus phase angle. COR2-A and COR2-B observations are colour coded in red and blue, respectively. Detections are shown as filled circles, whereas nondetections are shown as open circles. For comparison, we plot two models corresponding to light scattering of Phaethon dominated by its nucleus (dashed line, with the neighbouring shaded area representing the $\pm\!1\sigma$ uncertainty region) and a coma consisting of \micron-sized dust grains (black dotted line, with the $\pm\!1\sigma$ uncertainty region marked by two bracketing grey dotted lines). It is obvious that the obtained reduced magnitude measurements of Phaethon in the COR2 images are in line with the nucleus model but deviate considerably from the dusty coma model.
\label{fig:lc}
} 
\end{center} 
\end{figure}

The apparent magnitude of Phaethon is subject to varying observing geometry including heliocentric and observer-centric distances and phase angle. After normalising both distances to the mean Sun-Earth distance $r_{\oplus} = 1$ au, we obtained the reduced magnitude of Phaethon. Because our results were from image stacks combined from multiple individual exposures, during which the observing geometry varied nontrivially (see Table \ref{tab:vgeo}), we derived the relation between the apparent magnitude (including the $3\sigma$ thereof in case of nondetections) of Phaethon in the stacks ($\bar{m}$) and the corresponding reduced magnitude to be
\begin{equation}
%\nonumber
\bar{H}\left(\bar{\alpha} \right) %& = H\left(\bar{\alpha} \right)\\
 = \bar{m} + 2.5 \log \left(\frac{r_{\oplus}^{4}}{t_{2} - t_{1}} \int_{t_{1}}^{t_{2}}  \frac{{\rm d}t}{r_{\rm H}^{2} {\it \Delta}^{2}} \right)
\label{eq_mag_red}.
\end{equation}
\noindent Here, $t_1$ and $t_2$ are the start and end epochs of images used for stacking, $r_{\rm H}$ and $\it \Delta$ are respectively heliocentric and observer-centric distances both varying with time $t$, and the time-average phase angle $\bar{\alpha}$ was computed from
\begin{equation}
\bar{\alpha} = \frac{1}{t_{2} - t_{1}} \int_{t_{1}}^{t_{2}} \alpha \left(t \right) {\rm d}t
\label{eq_mean_PhA}.
\end{equation}
\noindent We plot the obtained reduced magnitude of Phaethon (see also the column labelled ``Red. Mag.'' in Table \ref{tab:phot}) as a function of phase angle in Figure \ref{fig:lc}, superimposed with two models assuming that light scattering of the asteroid is dominated by its nucleus and a dusty coma, respectively. For the nucleus model, we adopted the best-fit phase function of Phaethon's nucleus by \citet{2019AJ....158...30T}, which sampled phase angles from $\sim\!20$\degr~to 100\degr, and obtained the absolute magnitude of Phaethon's nucleus in the COR2 photometric system through transforming the absolute magnitudes of the nucleus in different bandpasses reported by \citet{2019AJ....158...30T} to the Pan-STARRS photometric system following \citet{2012ApJ...750...99T} and then applying the mean linear coefficients in Table \ref{tab:phot}. The $\pm1\sigma$ uncertainty region in the nucleus model is shown as the grey zone in Figure \ref{fig:lc}. For the dusty coma model, in accordance with the argument that the perihelion activity is associated with the ejection of \micron-sized dust grains based on HI-1 observations \citep{2013ApJ...771L..36J,2017AJ....153...23H}, we approximated the phase function of dust by the empirical Halley-Marcus model, which is applicable for cometary dust comparable to the transmitted wavelengths \citep{2007ICQ....29...39M,2011AJ....141..177S}. We still adopted the same absolute magnitude of Phaethon as in the nucleus model. The $\pm\!1\sigma$ uncertainty region in the dusty coma model is bounded by two grey dotted lines in Figure \ref{fig:lc}. We can immediately notice that the obtained reduced magnitude of Phaethon is consistent with the nucleus model, whereas the dusty coma model deviates significantly from the observations in particular at large phase angles $\ga\!150\degr$, where a strong forward-scattering enhancement is expected in the dusty coma model but is undoubtedly absent in the COR2 observations. 

Here, a few remarks are deserved to justify the adopted Halley-Marcus model for dust grains of Phaethon. Given the light scattering properties and size distribution of cometary dust \citep{2004come.book..565F,2004come.book..577K}, any ejected dust having sizes comparable to the transmitted wavelengths should dominate the overall signal of Phaethon's dusty coma, if it were present. Although the model is in good agreement with numerous comets all the way to even larger phase angles $\la \! 170\degr$ \citep{2007ICQ....29...39M,2013MNRAS.436.1564H}, it is also known that scattering of dust can be affected by various physical properties of dust, including the grain size, shape, porosity, and composition, and therefore discrepancies with the model may well occur. The major difference is in the forward-scattering regime as a result of diffraction, which is primarily dependent on the dust grain size \citep[and citations therein]{2004come.book..577K,2007ICQ....29...39M}. In reality, forward scattering of dust grains may be much more narrowly confined to higher phase angles \citep[e.g.,][]{2020ApJS..247...19M}, even beyond the coverage of the COR2 observations of Phaethon. However, this will require mm-sized or larger dust grains, not only order-of-magnitude greater than the transmitted wavelengths of the COR2 cameras, but also inconsistent with the argument that perihelion activity of Phaethon is associated with $\sim \! 1$ \micron-sized dust based on HI-1 observations \citep{2013ApJ...771L..36J,2017AJ....153...23H}. Furthermore, had a considerable amount of large dust grains been ejected from Phaethon, postperihelion observations would easily show a debris tail similar to the one of 323P/SOHO \citep{2022AJ....164...1H} but have not \citep{2005ApJ...624.1093H}, thereby suggesting the insignificance of large dust grains in the ejecta of Phaethon around perihelion. Given these, we think that our choice of the empirical Halley-Marcus model is appropriate for the putative dust grains of Phaethon. Therefore, we can confidently conclude that the observed behaviour of Phaethon in the COR2 data is consistent with the contribution solely from a bare nucleus.

\subsection{Activity}
\label{ssec_act}

Although the direct visual comparison between the COR2 observations of Phaethon and the two light-scattering models shows that the nucleus model is far better in agreement with the observations than the dusty coma model, it does not render us a quantitative extent to which the amount of dust can be ruled out. Thus, we proceed to evaluate the total effective scattering cross-section of \micron-sized dust grains that would be needed to account for the observed brightness of Phaethon in the COR2 data. The bandpasses of the COR2 cameras also guarantee that the observed signal is basically from dust scattering of sunlight alone and minimise possible contamination from gaseous emission lines that are typically seen for comets. 

Assuming the dusty coma exists, the total brightness of Phaethon consists of the contribution from its nucleus and the counterpart from the coma consisting of ejected dust grains. As we performed measurements in image stacks combined from individual exposures taken at different times, the results always refer to time-average values. We derived that the total effective scattering cross-section of ejected dust are related to the measured apparent magnitude of Phaethon by
\begin{equation}
\bar{{\it \Xi}}_{\rm d} = \frac{\pi r_{\oplus}^{2}}{p} 10^{0.4 \left(m_{\odot} - H_{\rm n}\right)} \left[\frac{t_2 - t_1}{r_{\oplus}^4} 10^{0.4 \left(H_{\rm n} - \bar{m} \right)} - \int_{t_1}^{t_2} \frac{\phi_{n}\left(\alpha\right)}{r_{\rm H}^2 {\it \Delta}^2} {\rm d}t\right] \cdot \left[\int_{t_1}^{t_2} \frac{\phi_{\rm d}\left(\alpha\right)}{r_{\rm H}^2 {\it \Delta}^2} {\rm d}t\right]^{-1}
\label{eq_Xd_avg}.
\end{equation}
\noindent Here, $\phi_{\rm n}$ is the phase function of the nucleus reported by \citet{2019AJ....158...30T}, $\phi_{\rm d}$ is the empirical Halley-Marcus model, $m_{\odot}$ is the apparent magnitude of the Sun in the COR2 photometric system at $r_{\oplus} = 1$ au, which we obtained using the {\it g}, {\it r}, and {\it i}-band apparent magnitudes of the Sun in the Pan-STARRS photometric system by \citet{2018ApJS..236...47W} and the mean linear coefficients listed in Table \ref{tab:phot}, and $p$ and $H_{\rm n}$ are respectively the geometric albedo and the absolute magnitude of Phaethon's nucleus, the former of which was adopted according to \citet{2019AJ....158...97M}. The uncertainty in $\bar{{\it \Xi}}_{\rm d}$ was properly propagated from errors in our photometric measurements, the adopted thermophysical parameters, and the nucleus model. 

We append the obtained values of $\bar{{\it \Xi}}_{\rm d}$ of Phaethon to Table \ref{tab:phot}, in the column of ``Cross-section", where one can immediately notice that they are all consistent with zero. This implies that no coma comprising \micron-sized dust is needed to account for the observed brightness of Phaethon in the COR2 images, and that the contribution from the nucleus alone is sufficient, in agreement with our conclusion from the visual comparison between the observations and the two models in Figure \ref{fig:lc}. The most stringent constraints on the amount of dust ejected around perihelion are provided by the nondetection observations taken at large phase angles ($\alpha \ga 150\degr$) due to forward scattering, which would enhance the intensity of dust over ten times in comparison to other illumination geometry. On the other hand, the nucleus would only have $\la \! 1\permil$ of the brightness at small phase angles, making its contribution (associated with the second term in the first square brackets in Equation \ref{eq_Xd_avg}) completely negligible. For instance, the observations of Phaethon in the apparition of 2009 from COR2-A and COR2-B were only a few days apart yet showed drastically different brightness. While the COR2-A observation at small phase angles implies a total effective scattering cross-section of $\left(5.9 \pm 34.8\right) \times 10^{6}$ m$^{2}$ for the ejected dust grains, corresponding to a $3\sigma$ upper limit of $\sim \! 10^{8}$ m$^{2}$, the COR2-B nondetection at large phase angles infers a much tighter $3\sigma$ limit of $\la \! 10^{5}$ m$^{2}$. 

Assuming that the ejected dust grains are spherical, with a mean radius of $\bar{\mathfrak{a}}_{\rm d}$ and a bulk density of $\rho_{\rm d}$, we can estimate the total ejected mass from the total effective scattering cross-section of dust through
\begin{equation}
\bar{M}_{\rm d} = \frac{4}{3}\rho_{\rm d} \bar{\mathfrak{a}}_{\rm d} \bar{{\it \Xi}}_{\rm d}
\label{eq_mloss}.
\end{equation}
\noindent Substituting with $\bar{\mathfrak{a}}_{\rm d} \sim 1$ \micron~and $\rho_{\rm d} = 2.6$ g cm$^{-3}$ \citep{2010IAUS..263..218B}, we obtain that the $3\sigma$ upper limit to the mass of ejected dust during a perihelion passage from the nondetections at large phase angles is $\la\!3 \times 10^{2}$ kg. In comparison, based on the HI-1 observation of Phaethon in the apparition of 2009, \citet{2013ApJ...771L..36J} reported a combined mass of $\sim\!3 \times 10^{5}$ kg, which is at least three orders of magnitude greater than the value permitted by the contemporary COR2-B observation. Other nondetections of Phaethon in the COR2 data at large phase angles from other apparitions cannot be explained by the coincidence that the perihelion brightening events all occurred outside the FOVs of the cameras. While indeed the nondetection observations from 2019 and 2022 only covered the preperihelion leg of Phaethon's orbit, the observation from 2013 also covered the timeframe in which the anomalous brightening would be expected to take place, given that perihelion activity of Phaethon was observed to be alike from apparition to apparition in HI-1 images \citep{2010AJ....140.1519J,2013AJ....145..154L,2017AJ....153...23H}. Thus, it seems highly unlikely that the perihelion activity of Phaethon is associated with the ejection of \micron-sized dust particles.

The COR2 observations do not negate the possibility that ejected dust particles having light scattering properties similar to the nucleus of Phaethon existed. However, such particles must be mm-sized or even larger debris to behave like mini-Phaethons. We can then use the COR2 observations at small phase angles to place a tighter constraint on the amount of debris, which may be shed around perihelion in a manner similar to near-Sun object 323P/SOHO \citep{2022AJ....164...1H}, albeit less stringent than the constraint for \micron-sized dust grains. The total effective scattering cross-section can be calculated from Equation (\ref{eq_Xd_avg}) by replacing $\phi_{\rm d}$ with $\phi_{\rm n}$ therein. What we found is that, the upper limit to the effective cross-section of the debris at small phase angles would be $\sim\!1.7 \pm 0.1$ times larger than the one for \micron-sized dust, i.e., $\bar{{\it \Xi}}_{\rm d} \la 10^8$ m$^2$ at $3\sigma$ level. By contrast, postperihelion observations of 323P were able to reveal its debris tail consisting of mm-sized and larger dust grains with a total effective scattering cross-section of $\ga \! 10^5$ m$^2$ within two months after the perihelion passage in 2021 \citep{2022AJ....164...1H}. We are unaware of any published literature that reported attempts to search for such a debris tail of Phaethon in a similar postperihelion timeframe. The closest was the attempt by \citet{2005ApJ...624.1093H}, who observed Phaethon about three months after the perihelion passage in 2003 and accordingly estimated an upper limit to the total effective scattering cross-section of dust to be $\la 10^5$ m$^2$. Yet given the fact that large debris would be relatively insusceptible to solar radiation pressure, we posit that the COR2 observations at small phase angles do not set a constraint on the debris tail of Phaethon better than the one based on ground-based observation by \citet{2005ApJ...624.1093H}. The fact that the debris tail of Phaethon is as yet undetected likely implies that large dust grains play a negligible role in the perihelion activity of the object, and/or that the ejection of such debris is a rare event with the occurrence rate of once in every $\ga$40 years, given the discovery time of Phaethon. It also remains unclear whether Phaethon has been contributing to the observed dust trail closely following the orbit of the asteroid reported by \citet{2020ApJS..246...64B,2022ApJ...936...81B} by ejecting large dust, and how. Nevertheless, we encourage future observations of Phaethon to be conducted using facilities far more sensitive than the COR2 cameras soon after perihelion so as to better search for the debris tail that might resemble the one of 323P/SOHO when opportunities arise (see Section \ref{sec_conj}).

\section{Conjecture}
\label{sec_conj}

\begin{figure}
\epsscale{0.8}
\begin{center}
\plotone{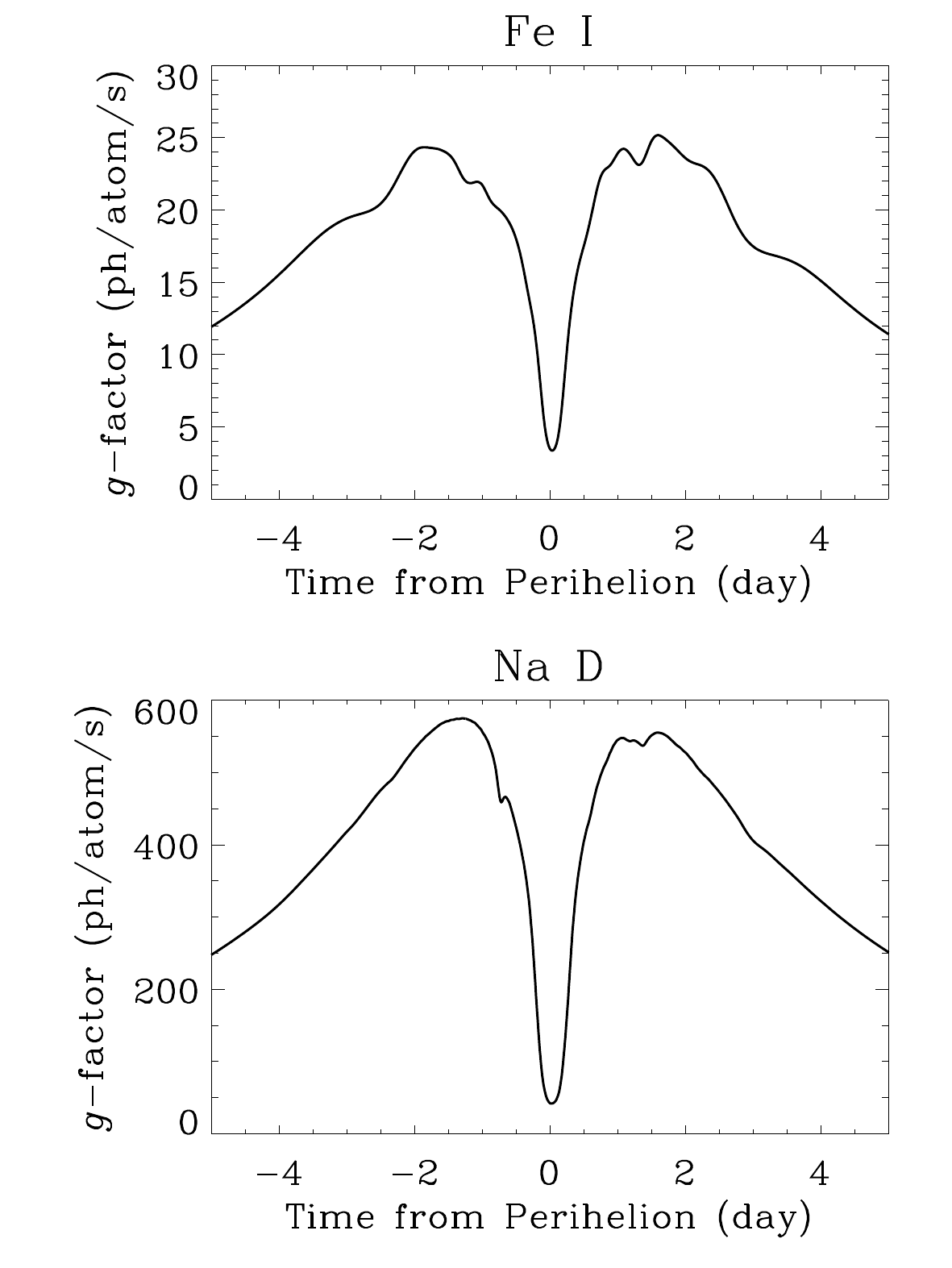}
\caption{
Fluorescence efficiencies ($g$-factors, in the unit of photons per atom per second) of Fe I (upper panel) and Na D (lower panel) emissions in the current orbit of Phaethon around perihelion. The dips at perihelion are due to the overlaps with the solar Fraunhofer absorption lines.
\label{fig:gfac}
} 
\end{center} 
\end{figure}

The COR2 observations of Phaethon clearly render the hypothesis that the observed perihelion activity at the asteroid is associated with the ejection of dust problematic. On one hand, the observed light scattering of Phaethon in the COR2 data severely contradicts the behaviour of \micron-sized dust grains because of the absence of forward scattering at large phase angles. On the other hand, the observations do not invalidate the existence of a dust environment around the nucleus of Phaethon dominated by mm-sized or larger dust grains. However, postperihelion observations of Phaethon should then easily detect a debris tail as in the case of 323P/SOHO \citep{2022AJ....164...1H} but did not. Furthermore, this type of dust grains are completely incompatible with the \micron-sized ones inferred from the HI-1 observations. Therefore, we can confidently conclude that the observed perihelion activity of Phaethon is highly unlikely relevant to the ejection of dust grains. So a natural question to ask is what is responsible for the perihelion activity of the asteroid in HI-1 data. We deduce that the signal is most likely from some fluorescence emissions transmittable to the HI-1 but not COR2 cameras onboard the STEREO spacecraft.

\begin{figure}
\epsscale{0.8}
\begin{center}
\plotone{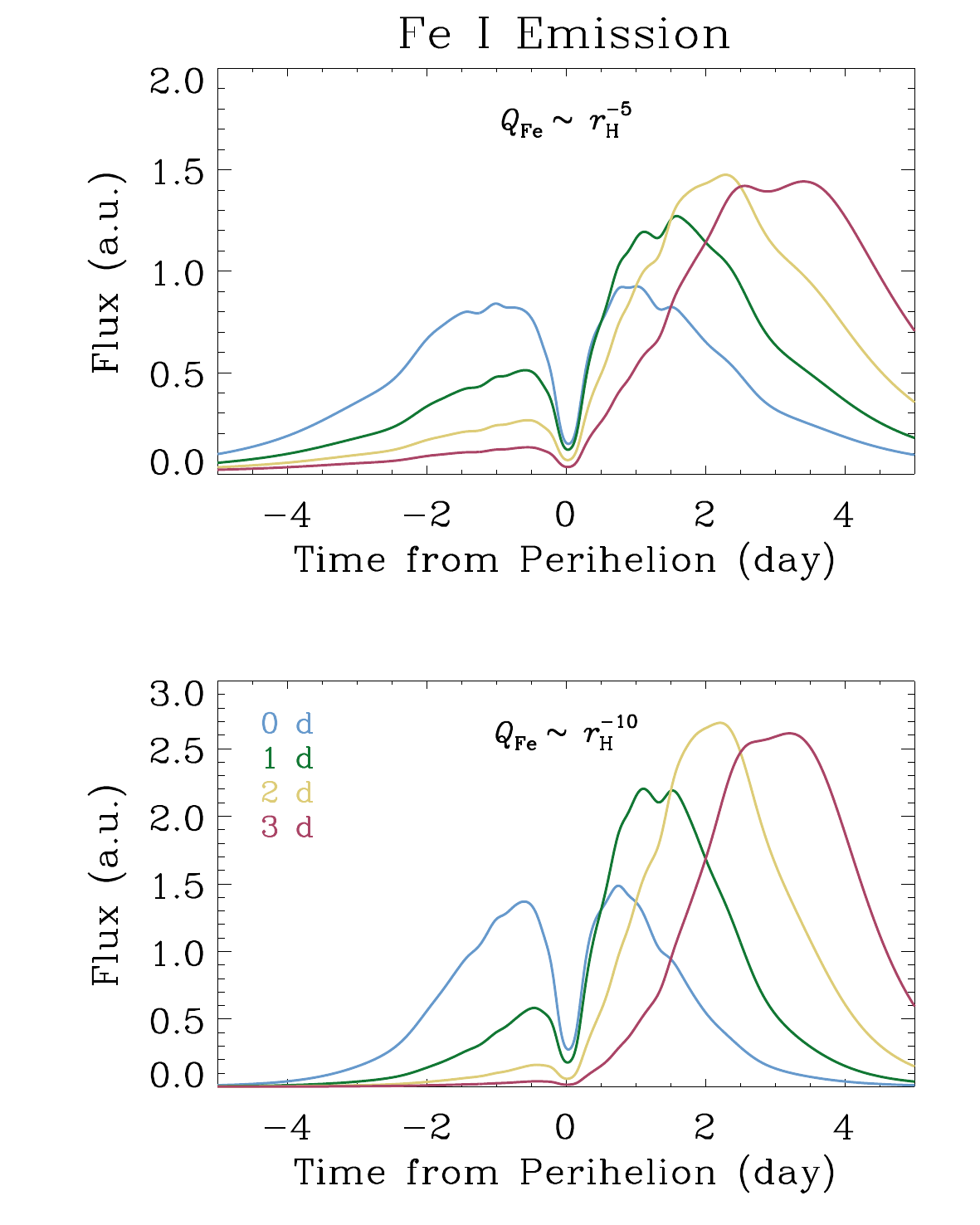}
\caption{
Modelled flux (in arbitrary units) of Phaethon due to Fe I emissions transmittable in the HI-1 cameras (wavelengths $\lambda \approx 400$ nm). In the top panel, the production rate of iron is assumed to be $\sim\! r_{\rm H}^{-5}$, whereas $\sim\! r_{\rm H}^{-10}$ is assumed in the bottom panel. Different colours of the curves correspond to the peak of the iron production rate delayed by different amounts of time (labelled in the legend) from perihelion.
\label{fig:model_Fe}
} 
\end{center} 
\end{figure}

\begin{figure}
\epsscale{0.8}
\begin{center}
\plotone{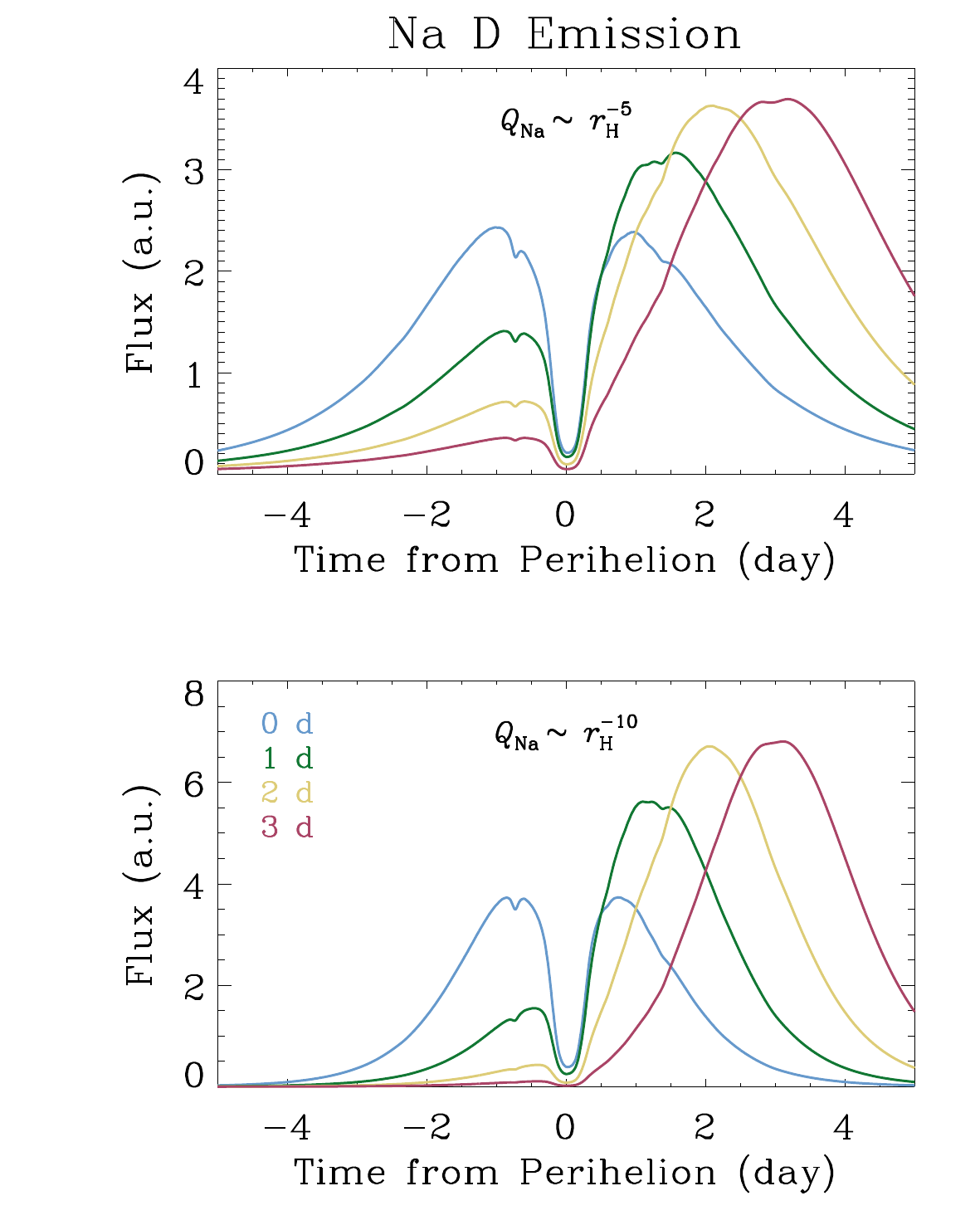}
\caption{
Same as Figure \ref{fig:model_Fe}, but for Na D-line emissions.
\label{fig:model_Na}
} 
\end{center} 
\end{figure}

The bandpasses of the HI-1 and COR2 cameras largely overlap with each other. However, there are still two noticeable differences: 1) the main bandpasses of the HI-1 cameras cover systematically shorter wavelengths, and 2) the HI-1 cameras have a nontrivial blue leak in their bandpasses around $\sim$400 nm \citep{2010SoPh..264..433B}. Therefore, we speculate that strong fluorescence emissions produced during the perihelion activity of Phaethon lies within either of the regions, or both. Associated with previous observations of comets such as C/1965 S1 (Ikeya-Seki) and C/2006 P1 (McNaught) at similar heliocentric distances \citep{1967ApJ...147..718P,2007ApJ...661L..93F}, and given the study by \citet{2021Natur.593..372M} finding that free iron atoms prevail in comae of comets even at larger heliocentric distances, we boldly conjecture the Fe I emission lines around $\lambda \approx 400$ nm being a promising possibility. The conjecture does not conflict with the model prediction by \citet{2022Icar..38114995L} that Phaethon can develop a coma comprised of iron gas around perihelion. Future observations of the asteroid are needed to verify the thermophysical model by \citet{2022Icar..38114995L} and our conjecture.

Besides, we are aware that Mercury was observed to possess an anti-sunward tail in images taken by the HI-1 cameras on multiple occasions \citep{2013PhDT.......602S}. Given the effective transmission of the HI-1 cameras, the tail appeared to be too bright to be accounted for by sodium D emissions, unless degradation has occurred to the HI-1 cameras that caused a change in their bandpasses, making sodium D emission lines transmittable \citep{2013PhDT.......602S}. Recently, \citet{2022Icar..389..115218P} reported the identification of a prominent neutral sodium tail at comet C/2011 L4 (PANSTARRS) in HI-1 observations, whereby they also suspected a considerable shift in the bandpass of the camera. Unfortunately, we are unaware of any published literature on measuring the ageing effects of the HI-1 filters. Instead, the focus was primarily on the detector sensitivity, which was reported to have little sign of degradation \citep{2012SoPh..276..491B}. Although the STEREO team did notice a shift in the bandpasses, relevant details are yet to be characterised (K. Battams, private communication). More recently, Zhang et al. (submitted) analysed observations of Phaethon around its latest perihelion return in 2022 May from the SOHO spacecraft \citep{1995SoPh..162....1D}, finding that it was only visible through the orange filter. The markedly higher intensities of near-Sun comets in the orange filters are interpreted to be caused by the sodium D emissions \citep{2010AJ....139..926K}. By the same token, Zhang et al. suggested to us that the sodium emissions likely dominated the observed signal of Phaethon in the SOHO images. In addition, they analysed earlier SOHO and HI-1 observations, arriving at the same conclusion that the sodium emissions most likely played a crucial role in causing the observed anomalous brightening of the object around perihelion. Taking into account the fact that sodium emissions have been observed in many comets at heliocentric distances $\la$0.7 au, the length and brightness of sodium tails of comets are found to climax at heliocentric distances in a range of $\sim\!0.1$-0.2 au \citep{1970A&A.....5..286H}, and the results from the thermophysical modelling and laboratory experiments by \citet{2021PSJ.....2..165M} that sodium emissions around perihelion passages of Phaethon are capable of producing the observed activity, here we also include sodium emissions to form a comparison to the results with iron emissions. 

In the following, we apply a simplistic model to briefly assess whether Fe I and/or sodium D emissions can qualitatively reproduce the observed flareup in the brightness of Phaethon around its perihelion passages \citep{2010AJ....140.1519J,2013AJ....145..154L,2017AJ....153...23H}. A more sophisticated model for application of the sodium emissions will be separately presented by Zhang et al. (submitted). First, we need to compute the fluorescence efficiency, or called the $g$-factor, which essentially describes the rate at which an atomic species absorbs photons, from \citep[c.f.][]{1979ApJ...234.1148S}
\begin{equation}
g \left(r_{\rm H}, \dot{r}_{\rm H} \right) = \gamma\left(\dot{r}_{\rm H} \right) \frac{q_{\rm e}^2 f \lambda^3 F_{\lambda}}{8 \varepsilon_0 \hbar \mathfrak{m}_{\rm e} c^3} \left(\frac{r_{\oplus}}{r_{\rm H}}\right)^2
\label{eq_gfac}.
\end{equation}
\noindent Here, the fluorescence efficiency is a function of both the heliocentric distance and the rate thereof, $\gamma$ is the fraction of the Doppler-shifted solar incident flux relative to the solar continuum, whose spectral radiance per unit wavelength is denoted as $F_{\lambda}$, $f$ is the absorption oscillator strength, $q_{\rm e} = 1.6 \times 10^{-19}$ C is the elementary charge, $\hbar = 1.1 \times 10^{-34}$ J s$^{-1}$ is the reduced Planck's constant, $\varepsilon_0 = 8.9 \times 10^{-12}$ F m$^{-1}$ is the vacuum permittivity, $\mathfrak{m}_{\rm e} = 9.1 \times 10^{-31}$ kg is the electron mass, and $c = 3.0 \times 10^{8}$ m s$^{-1}$ is the speed of light. Physical parameters for sodium D and Fe I emission lines were taken from \citet{Steck2010} and \citet{2003ApJS..149..205M}, respectively, and the high-resolution solar spectrum atlas by \citet{1984sfat.book.....K} was used in the calculation. The fluorescence efficiencies in the current orbit of Phaethon within five days from perihelion are plotted in Figure \ref{fig:gfac}.

Next, we assume various heliocentric dependencies for the atomic production rates $Q$ of iron and sodium atoms. In our simplistic model, the total flux due to the fluorescence emission is proportional to $g Q \tau$, where $\tau$ is the photoionisation lifetime scaled as the square of the heliocentric distance. We adopted the photoionisation lifetimes of iron and sodium based on \citet{2015P&SS..106...11H}. Given the HI-1 observations of the peak flux of Phaethon during the perihelion activity, we can then estimate the corresponding production rates of iron and sodium to be $Q_{\rm Fe} \sim 10^{25}$ s$^{-1}$ and $Q_{\rm Na} \sim 10^{24}$ s$^{-1}$, respectively. In comparison, \citet{2007ApJ...661L..93F} analysed also the HI-1 images and reported a production rate of $Q_{\rm Fe} \sim 10^{30}$ s$^{-1}$ for comet C/2006 P1 (McNaught) at heliocentric distance $r_{\rm H} = 0.25$ au. \citet{2021PSJ.....2..228B} and \citet{2021Natur.593..372M} measured the production rates of iron to be $Q_{\rm Fe} \sim 10^{21}$-10$^{24}$ s$^{-1}$ for a list of comets at heliocentric distances of $\sim$0.7-3 au. As for sodium, for instance, comet C/1995 O1 (Hale-Bopp) was measured to produce sodium atoms at a rate of $\sim\!10^{25}$-$10^{26}$ s$^{-1}$ at $\sim$1 au from the Sun \citep{1997ApJ...490L.199C,1998A&A...334L..61R,1998GeoRL..25..225W}, and C/2012 S1 (ISON) had $Q_{\rm Na} \sim 10^{23}$-10$^{24}$ s$^{-1}$ \citep{2015Icar..247..313S}. Here, with the comparisons we do not mean to suggest that sodium and iron are produced in a common manner at Phaethon and the aforementioned comets at larger heliocentric distances; the process occurring at Phaethon near perihelion is possibly confined to the near-Sun environment only, where the surface temperature can be heated up to $\ga \! 1000$ K.

The obtained general shapes of the emissions due to Fe I and Na D lines are similar, both of which brighten around perihelion but drop considerably at perihelion, because of the overlaps with the solar Fraunhofer absorption lines. In reality, the behaviour of the perihelion activity of Phaethon is observed to be repetitive in different perihelion passages, which peaks approximately a day after perihelion passage \citep{2010AJ....140.1519J,2013AJ....145..154L,2017AJ....153...23H}. We boldly conjecture that this is possibly a consequence of the thermophysical heterogeneity due to the spin-axis orientation of Phaethon \citep{2022arXiv220308865M}, whereby the northern hemisphere is possibly more thermally processed and the southern one is suddenly exposed to sunlight soon after perihelion \citep{2009PASJ...61.1375O,2014ApJ...793...50A,2019MNRAS.482.4243Y}. Therefore, we arbitrarily shift the peak of the production rate by 1-3 days past perihelion in the same fashion as in the asymmetric outgassing model for cometary nongravitational forces \citep{1989AJ.....98.1083Y}, and recomputed the emission fluxes. We show the results for Fe I and Na D emissions respectively in Figures \ref{fig:model_Fe} and \ref{fig:model_Na}, in which we can see the best similarity to the HI-1 photometry of Phaethon \citep{2010AJ....140.1519J,2013AJ....145..154L,2017AJ....153...23H} is given by the asymmetric models with a delay of one or two days in the peak production rate. 

Admittedly, at the current stage, by no means can we affirm that the observed perihelion activity of Phaethon is due to iron and/or sodium emissions, but we posit that the hypothesis is a promising one, in that it has no conspicuous conflict with the HI-1 and COR2 observations, whereas the dust hypothesis encounters insurmountable difficulties. In order to identify which specific gas species are the cause of the perihelion activity of Phaethon, the only feasible way is to observe the asteroid close to the Sun as much as possible, ideally at heliocentric distances $\la\!1$ au, given the fact that sodium emissions have been observed at many comets at $r_{\rm H} \la 1$ au, mostly $\la \! 0.7$ au \citep{1970A&A.....5..286H}. We expect that in-situ measurements from the upcoming DESTINY$^{+}$ mission \citep{2018LPI....49.2570A}, which is presently planned to have a flyby with Phaethon at a heliocentric distance of 0.87 au in 2026 August \citep{2019P&SS..172...22K}, may be able to test our gas hypothesis. 

To facilitate future ground observations of Phaethon, we exploited JPL Horizons to search for potential future ideal observing windows of Phaethon at heliocentric distances $r_{\rm H} \le 1$ au while at solar elongations $\ge\!50\degr$ by the end of this century. The opportunities in the next decade are tabulated in Table \ref{tab:vgeo_future}. These will still be applicable for telescopes that are capable of reaching lower solar elongations, just that the observing windows will be wider. Given the current orbit, unfortunately Phaethon can never be observed at $r_{\rm H} \le 0.7$ au and solar elongations $\ga \! 50\degr$ simultaneously from the ground in the investigated period of time. However, we managed to identify a number of observing windows of Phaethon at even smaller heliocentric distances during total solar eclipses by the end of the century, during which time searching for gas emissions of the asteroid will be feasible for ground observations (see also Table \ref{tab:vgeo_future}). The best of all opportunities will be during the total solar eclipse on 2089 October 4, soon after the perihelion passage of Phaethon, when the asteroid will be at a heliocentric distance of $\sim\!0.16$ au and $\sim\!8\degr$ from the Sun. Far more observations of Phaethon at small heliocentric distances will be certainly needed for us to fully understand the perihelion activity of the object.

\begin{deluxetable*}{lcccc}
%\tabletypesize{\scriptsize}
\tablecaption{Future Ideal Observing Windows of Phaethon
%\rotate
\label{tab:vgeo_future}}
\tablewidth{0pt}
\tablehead{
\colhead{Observable Date} & \colhead{Heliocentric Distance} & \colhead{Geocentric Distance} & \colhead{Solar Elongation} & \colhead{Phase Angle} \\
\colhead{(UT)} & \colhead{$r_{\rm H}$ (au)} & \colhead{${\it \Delta}$ (au)} & \colhead{$\varepsilon$ (\degr)} & \colhead{$\alpha$ (\degr)}
}
\startdata
2023 Nov 17 -- 2023 Nov 28 & 0.791$\,\to\,$0.992 & 0.788$\,\to\,$0.898 & 51.4$\,\to\,$63.3 & 77.6$\,\to\,$62.7 \\ 
2025 Feb 17 -- 2025 Feb 24 & 0.996$\,\to\,$0.873 & 1.083$\,\to\,$1.066 & 57.3$\,\to\,$50.1 & 56.5$\,\to\,$60.4 \\
Total Solar Eclipse on 2026 Aug 12 & 0.648 & 1.542 & 17.2 & 27.6 \\
2026 Oct 1 -- 2016 Oct 10\tablenotemark{$\dag$} & 0.823$\,\to\,$0.986 & 0.387$\,\to\,$0.368 & 52.2$\,\to\,$77.3 & 106.0$\,\to\,$81.3 \\
2027 Dec 31 -- 2028 Jan 11 & 1.000$\,\to\,$0.801 & 0.290$\,\to\,$0.379 & 84.8$\,\to\,$50.9 & 78.4$\,\to\,$107.5 \\
2029 Aug 12 -- 2029 Aug 23 & 0.798$\,\to\,$0.998 & 0.809$\,\to\,$0.824 & 50.4$\,\to\,$64.9 & 78.1$\,\to\,$66.6 \\
2030 Nov 13 -- 2030 Nov 24 & 0.990$\,\to\,$0.789 & 0.546$\,\to\,$0.465  & 74.0$\,\to\,$51.7 & 74.0$\,\to\,$100.8 \\
Total Solar Eclipse on 2030 Nov 25 & 0.763 & 0.466 &  48.5 & 104.2 \\
Total Solar Eclipse on 2078 May 11 & 0.813 & 1.672 &  20.9 &  26.3 \\
Total Solar Eclipse on 2082 Aug 24 & 0.694 & 0.632 &  42.6 &  99.3 \\
Total Solar Eclipse on 2088 Apr 21 & 0.250 & 0.961 &  14.4 &  92.9 \\
Total Solar Eclipse on 2089 Oct 4 & 0.158 & 1.049 &   8.4 &  67.9 \\
Total Solar Eclipse on 2094 Jan 16 & 0.215 & 0.832 &   9.7 & 129.8 \\
Total Solar Eclipse on 2095 Jun 2 & 0.764 & 1.755 &   8.2 &  10.8 \\
Total Solar Eclipse on 2099 Sep 14 & 0.863 & 1.385 &  38.3 &  46.3 \\
\enddata
\tablenotetext{\dag}{The perigee will occur approximately on 2026 Oct 8 at a distance of ${\it \Delta} \approx 0.365$ au.}
\tablecomments{The observing geometry of Phaethon during a total solar eclipse is referred to the corresponding time of greatest eclipse given at {\url{https://eclipse.gsfc.nasa.gov/SEcat5/SE2001-2100.html}}.}
\end{deluxetable*}

%In the absorption process, the atom experiences an acceleration by exchanging momentum with incident photons. The ratio between this radiation acceleration and the solar gravitational acceleration is then %\citep[c.f.][]{1997ApJ...490L.199C}
%\begin{equation}
%\beta = \frac{2 \pi \hbar r_{\rm H}^2}{GM_{\odot} \lambda \mathscr{U} \mathfrak{m}_{\rm H}} g \left(r_{\rm H}, \dot{r}_{\rm H} \right)
%\label{eq_beta},
%\end{equation}
%\noindent where $G = 6.7 \times 10^{-11}$ m$^3$ kg$^{-1}$ s$^{-2}$ is the gravitational constant, $M_{\odot} = 2.0 \times 10^{30}$ kg is the mass of the Sun, $\mathfrak{m}_{\rm H} = 1.7 \times 10^{-27}$ kg is the mass of the hydrogen atom, and $\mathscr{U}$ is the atomic mass of the species.

%\clearpage
\section{Summary}
\label{sec_sum}

We studied near-Sun asteroid (3200) Phaethon using coronagraphic observations taken by the COR2 cameras onboard the STEREO spacecraft around different perihelion returns. The key results are:

\begin{enumerate}

\item Although Phaethon is invisible in individual COR2 images, we managed to spot it in image stacks combined from the same perihelion observations. However, the asteroid is visible only at small ($\la$30\degr) but not large phase angles ($\ga$150\degr). The observed results are consistent with light scattering dominated by the nucleus of Phaethon, rather than by a coma comprising \micron-sized dust grains, the latter of which was inferred to exist based on HI-1 observations by \citet{2013ApJ...771L..36J} and \citet{2017AJ....153...23H}.

\item Using the nondetection observations at large phase angles, we obtained that the total effective scattering cross-section of \micron-sized dust grains ejected during a perihelion passage is $\la \! 10^5$ m$^{2}$ at $3\sigma$ level, at least three orders of magnitude smaller than previous estimates based on HI-1 observations alone. Thereby the argument that perihelion activity of Phaethon is associated with \micron-sized dust is called into serious question.

\item The COR2 observations cannot rule out the existence of mm-sized or larger debris dominating the dust environment around Phaethon that behave like mini-Phaethons. However, the nondetection of a postperihelion debris tail suggests the negligible role of such dust in the perihelion activity of the asteroid.

\item We thereby conclude that perihelion activity of Phaethon is highly unlikely relevant to the ejection of dust grains. Rather, we conjecture that the activity is possibly accounted for by Fe I and/or Na D emission lines, the latter of which may have become transmittable to the HI-1 cameras because of the ageing effect.

\item We modelled the fluxes of Phaethon due to Fe I and Na D emissions, finding that the asymmetric models in which the peak of the atomic production rate is delayed by $\sim$1 day from perihelion can best reproduce a lightcurve qualitatively similar to the HI-1 observations.

\item Our conjecture must be validated by future observations of Phaethon at small heliocentric distances dedicated to the search of its gas emissions. More observations are needed for us to fully understand the perihelion activity of the asteroid. We identified a list of observing windows ideal for future ground observations, the best of which will be during total solar eclipses.

\end{enumerate}

\begin{acknowledgements}
We thank Bin Yang, Karl Battams, and William Thompson for their help and Qicheng Zhang for discussions and two anonymous reviewers for providing insightful comments on our manuscript. We thank also Frederic A. Rasio spending time reviewing the manuscript and commenting that our work is of rather limited interest to the research community as a whole. This work was supported by the Science and Technology Development Fund, Macau SAR, through grant Nos. 0016/2022/A1 and SKL-LPS (MUST)-2021-2023. 
\end{acknowledgements}

\vspace{5mm}
\facilities{STEREO}

\software{{\tt astrometry.net} \citep{2010AJ....139.1782L}, {\tt IDL}, {\tt MPFIT} \citep{2009ASPC..411..251M}, {\tt SSW} \citep{1998SoPh..182..497F}}

\end{document}